\documentclass[osajnl,twocolumn,showpacs,superscriptaddress,11pt]{revtex4-1} 
\usepackage{amsmath,amssymb,graphicx}
\usepackage[english]{babel}
\usepackage[latin1]{inputenc}

\begin{document}
\title{Vanishing tilt-to-length coupling for a singular case in two-beam laser interferometers with Gaussian beams}
 
\author{Sönke Schuster}\email{Corresponding author: soenke.schuster@aei.mpg.de}
\author{Gudrun Wanner}
\author{Michael Tröbs}
\author{Gerhard Heinzel}

\affiliation{
Max Planck Institute for Gravitational Physics (Albert Einstein Institute) and Institute for Gravitational
Physics of the Leibniz Universität Hannover, Callinstr. 38, D-30167 Hannover, Germany
}

\begin{abstract} 
The omnipresent tilt-to-length coupling in two-beam laser interferometers, frequently a nuisance in precision measurements, vanishes for the singular case of two beams with identical parameters and complete detection of both beams without clipping. This effect has been observed numerically  and is explained in this manuscript by the cancellation of two very different effects of equal magnitude and opposite sign.
\end{abstract}

\maketitle

\section{Introduction}
One recurring noise source in precision interferometric length measurements is the parasitic coupling of misalignments (tilt) into the length readout, which arises due to straightforward geometrical pathlength changes of the beam axis.
In this manuscript we show that in the special case of two identical fundamental Gaussian beams, a large detector without any clipping and a pivot placed directly on the beam axis another effect of the same magnitude and the opposite sign occurs which to first order cancels the geometrical pathlength change in the interferometric measurement.

This manuscript does not investigate the effect of a lateral offset between pivot and beam axis, but only the effect of a lever arm between pivot and photodetector.

\section{Geometrical coupling}
\label{sec: analy computation}
\begin{figure}[h]
\centerline{\includegraphics[width= 0.4\columnwidth]{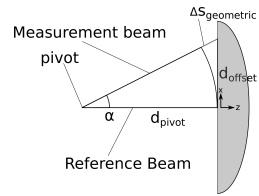}}
\caption{The reference beam is centered on the detector the measurement beam is tilted by the angle $\alpha$ around the pivot. The geometric pathlength change $\Delta$s is the additional distance the measurement beam has to propagate until it reaches the detector.}
\label{fig: geom anschaulich}
\end{figure}

We consider a simplified interferometer reduced to  its essential components.
Only the reference beam, the (tilted) measurement beam, and the photodiode are considered (Fig.~\ref{fig: geom anschaulich}). 
The photodiode detects the interference pattern between the measurement and reference beams,  and from its photo-current it is possible to determine the phase difference  between the two beams by a variety of different readout schemes, both homodyne and heterodyne~\cite{Wanner2012}.
That phase difference can be translated to the longitudinal pathlength sensing signal ($s_{\mathrm{LPS}}$) that describes the difference in the travelled pathlength between the two beams~\cite{Wanner2012}. The result is independent of the interferometer type (homodyne or heterodyne).  
Any tilt of the measurement beam causes a tilt-to-length coupling.

Intuitively, one expects a coupling between the measurement beam angle and the pathlength change: The beam tilt results in a longer distance that the measurement beam has to travel from the pivot to the photodiode. Using straightforward geometry, this geometric pathlength change $\Delta s_\text{geo}$  can be computed analytically:
\begin{equation}
\Delta s_\text{geo} = \left(\frac{1}{\cos(\alpha)}-1 \right) d_\text{pivot} \approx \frac{\alpha^2}{2} d_\text{pivot} + \mathcal{O}(\alpha^4)\,.
\label{eq: geometric}
\end{equation}
Here, $\alpha$ is the beam angle and $d_\text{pivot}$ the distance between pivot and photodiode (as shown in Fig.~\ref{fig: geom anschaulich}).
One would expect that  this geometric pathlength change always appears in the measured $s_{\mathrm{LPS}}$. This manuscript shows that this is indeed true for plane waves, but not for Gaussian beams. 

\section{Plane Waves}
\label{sec: plane wave}
In this section, the relation between the geometrical pathlength change and the $s_{\mathrm{LPS}}$ is discussed for the case of two plane waves.
The electric field for an infinite plane wave is given by:
\begin{equation}
E_\text{plane}(x,y,z) = A \, \exp\left( -i \omega t -i k z  + i\it{\Phi} \right) \, ,
\end{equation}
if the plane wave propagates in $z$ direction, where $k=2\pi /\lambda$ is the wave number, $\omega$  the frequency, $A$  the amplitude and $\it{\Phi}$ the initial phase.   
This expression is used for the reference beam $E_\text{ref} (x,y,z) = E_\text{plane}(x,y,z)$ and a rotated version is used for the tilted measurement beam.
We denote the coordinate system of $E_\text{ref}$ by $\vec{\mathbf{r}}_\text{ref}$,
the location of the pivot by $\vec{\mathbf{p}}_\text{pivot}$,
the rotation-matrix for a rotation around the y-axis is $\hat{\mathbf{m}}_\text{rot}$,
and the resulting coordinate system of the second electric field $E_\text{meas}$ is called $\vec{\mathbf{r}}_\text{meas}$:

\begin{equation*}
\vec{\mathbf{r}}_\text{ref} = \begin{pmatrix} x \\ y \\ z  \end{pmatrix}\, , \quad 
\vec{\mathbf{p}}_{\text{pivot}} = \begin{pmatrix} 0 \\ 0 \\ -d_\text{pivot}  \end{pmatrix} \, ,\quad 
\end{equation*}
\begin{equation}
\hat{\mathbf{m}}_\text{rot} = \begin{pmatrix} \cos \alpha& 0 & -\sin \alpha \\ 0 & 1 & 0 \\ \sin \alpha & 0 & \cos \alpha   \end{pmatrix}
\end{equation}
\begin{equation}
\vec{\mathbf{r}}_\text{meas} = \hat{\mathbf{m}}_\text{rot}^{-1} \cdot \left( \vec{\mathbf{r}}_\text{ref} - \vec{\mathbf{p}}_{\text{pivot}}\right) + \vec{\mathbf{p}}_{\text{pivot}} \, .
\label{eq: rotation}
\end{equation}
A more detailed explanation of this transformation can be found in~\cite{Wanner2014}.
The tilted electric field is now defined similar to $E_\text{ref}$, but with new coordinates:
\begin{equation}
E_\text{meas}(x,y,z)=E_\text{plane}(\vec{\mathbf{r}}_\text{meas}(x,y,z)) \, .
\end{equation}
Since the $z$ position of the photodiode plane is arbitrary, it can be set to zero. 
The pathlength difference between the two beams is encoded in the intensity of the superposition between the two beams 
and thus also in the power as measured by a photodiode.  
By computing the integral of the intensity over the entire sensitive area, it is possible to extract the pathlength difference by analysing the power fluctuation on the photodiode~\cite{Wanner2012}. 
The same information is also covered in the complex phase of the integral over the overlap term: 
\begin{align}
\arg\left(\int_{\text{pd}} E_\text{meas}E_\text{ref}^* \,\mathrm{d}r^2\right) &=  ks_\text{LPS} \, .
\label{eq: weighting}
\end{align}
We prefer to extract the phase from the complex overlap term (Eq.~(\ref{eq: weighting})) instead of from the power variation as this reduces the computational effort.
Since the $s_{\mathrm{LPS}}$ does not change in time and we are only interested in the variation of the phase difference between the two beams,  we can set $t = 0$ and the initial total phase  $\Phi = 0$.
An integration of the overlap term over a square detector at position $z=0$ (side length $2 r_\text{pd}$) gives the overlap integral for plane waves $O_\text{ovi}^{P}$, which corresponds to the complex amplitude in~\cite{Wanner2012}:
\newline

\begin{widetext}
\begin{align}
O_\text{ovi}^{P}  =&  A_\text{ref} A_\text{meas}\frac{4 r_\text{pd} \left\{ \cos\left[k d_\text{pivot} \left(-1+\cos\alpha\right)\right]-i \sin(-k d_\text{pivot}+k d_\text{pivot} \cos\alpha) \right\} \sin(k d_\text{pivot} \sin\alpha)}{k \sin\alpha} \, .
\end{align}
\end{widetext}
The complex phase of this integral describes the phase difference between the two plane waves. This phase difference can be translated to the $s_{\mathrm{LPS}}$ using the wave number $k$:
\begin{equation}
s_\text{LPS} = \frac{\arg (O_\text{ovi}^{P})}{k}   \approx \frac{\alpha^2}{2} d_\text{pivot} + \mathcal{O}(\alpha^4) \approx \Delta s_\text{geo} \, . \label{eq: result plane wave}
\end{equation}
Thus, two plane waves on a detector show approximately the geometrical coupling $\Delta s_\text{geo}$, confirming the intuitive results from Eq.~(\ref{eq: geometric}).

\section{Gaussian Beams}
\label{sec: gaussian beam}
In laser interferometers, fundamental Gaussian beams are a more appropriate description than plane waves. We start with the special case of two identical fundamental Gaussian beams and an infinite detector (i.e. both beams are completely detected without any clipping).
The amplitude of the electric field is irrelevant for the pathlength signal and is therefore set to unity. The Gouy phase is also ignored, 
since its offset is negligible in the case of equal beams.
The electric field can then be written as  \cite{Yariv1989,Saleh1991}:
\begin{equation}
E_\text{Gauss}(x,y,z)= \exp\left(-i\omega t -i k \frac{x^2 + y^2}{2 q} - i k (z-z_0)\right) \, ,
\label{eq: beam}
\end{equation}
with the complex $q$ parameter $q=(z - z_0)+iz_r$,
where $z_r$ is the Rayleigh range and $z_0$ denotes the position of the waist. The term $-ik(z-z_0)$ can be set to $-ikz$ since the $z_0$ dependent phase shift will not change. 
The expression in Eq.~(\ref{eq: beam}) is used for the reference beam $E_\text{ref} (x,y,z)$, and 
a rotated version  is used for the tilted measurement beam.
The coordinate system of the tilted beam is computed analogously to Eq.~(\ref{eq: rotation}).
The real part of the $q$ parameter changes only by propagation in beam direction. For the reference beam the propagation corresponds to an increase in $z$. 
For the measurement beam, the change in  direction of propagation corresponds to an increase in $z$ but also a change of $x$ (Eq.~(\ref{eq: rotation})).  
This $x$ dependence makes the 2D integration in the detector plane much harder.
Therefore the real part of the $q$ parameter $z-z_0$ is set to the constant value $-z_0$. 
The changes in  $z$  due to the coordinate transformation over the detector surface are very small and cause a negligible changing (therefore the error produced by a $z$-independent $q$ parameter becomes also very small and is neglected).
Furthermore an infinite detector is assumed, which practically means any single element photodiode (SEPD) that is larger than three times the beam size. The integral of the overlap term over an infinite detector at position $z=0$ yields: 
\newline
\begin{widetext}
\begin{equation}
\begin{split}
O_\text{ovi}^{G} =&  \frac{{2}  \pi  \left(z_0^2+z_r^2\right)}{k \sqrt{z_r \left[-i z_0+3z_r+(z_r+i z_0) \cos2\alpha\right]}} \exp\left[\frac{2 i k \xi\sin\left(\alpha/2\right)^2}{-z_0-3 i z_r+(z_0-i z_r) \cos (2 \alpha)}\right]
 \end{split}
\end{equation}
with:
\[
\xi = (z_0+d_\text{pivot})^2+2 i d_\text{pivot} z_r+z_r^2+\left[(z_0+d_\text{pivot})^2-2 i d_\text{pivot} z_r+z_r^2\right] \cos\alpha\,.
\]
\end{widetext}
This leads to the resulting pathlength change:
\begin{equation}
s_\text{LPS} = \frac{\arg (O_\text{ovi}^{G} )}{k}  \approx 	\frac{\alpha^2 z_0}{4 z_r k} + \mathcal{O}(\alpha^4) \approx 0 \, .
\label{eq: phase to LPS}
\end{equation}
This result matches the expressions in~\cite{Wanner2014} for the special case of equal beams.
For two plane waves the resulting coupling (Eq.~(\ref{eq: result plane wave})) 
has the same form (proportional to $\alpha^2$) with a prop. factor given by $d_\text{pivot}/2$, which usually is a macroscopic quantity of magnitude between centimeters and meters.
For two Gaussian beams, this factor becomes $z_0 /(4z_r k)$ which is 
of the same order of magnitude as the wavelength i.e. nanometers to micrometers for visible or infrared light.
For typical parameters and beam angles $\approx 1\,$mrad, the resulting length change is significantly below pico meter scale and thus below the sensitivity of most interferometers.

\begin{center}
\begin{figure}
\centerline{
\includegraphics[width= 0.9\columnwidth]{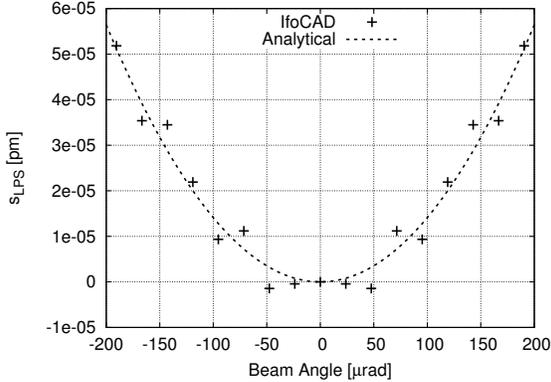}}
\caption{Numerically/Analytically computed $s_{\mathrm{LPS}}$ (second order approximation) for  Gaussian beams. The differences are due to numerical errors.}%
\label{fig: GG}%
\end{figure}
\end{center}

All results in this manuscript were confirmed by  numerical simulations computed with IfoCAD \cite{Wanner2012,Kochkina2012}.
Exemplary results for the actual setup are shown in Fig.~\ref{fig: GG}. For a wavelength of 1064\,nm, waist radii of 1\,mm,
30\,mm photodiode diameter and the pivot and waists located 100\,mm in front of the photodiode. 

\begin{figure}
\centerline{\includegraphics[width=0.6 \columnwidth]{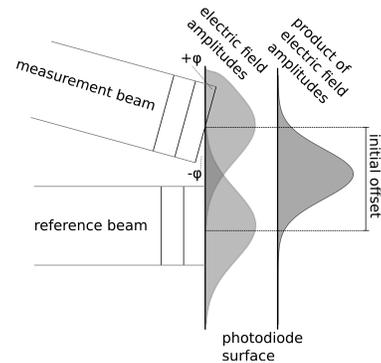}}
\caption{The reference beam is centered on the SEPD, the measurement beam has an offset and is tilted around its center. This offset results in a different weighting of the phase differences, the negative phase part has a higher weighting and the resulting coupling shows a negative phase.}
\label{fig: off anschaulich}
\end{figure}

In the remaining part of the manuscript, we will show that the reason for the vanishing coupling for two Gaussian beams is an additional coupling effect which is caused by an angle depending offset.
A beam tilt generates, besides the geometric change of the pathlength, two other effects.
The first one is a relative angle between the two beams on the photodiode and the other is an offset between them ($d_\text{offset}$ in Fig.~\ref{fig: geom anschaulich}). 
Since infinite plane waves have no uniquely defined center, any shift orthogonal to their direction of propagation maps the wave upon itself and causes therefore no effect.
This is different for Gaussian beams: Due to the Gaussian intensity profile, there is a uniquely defined center. 
To investigate the effect of the generated offset in the case of Gaussian beams, the initial setup (Fig.~\ref{fig: geom anschaulich}) is changed
to create a situation with an angle-invariant offset and no lever arm. We place the pivot directly on the detector. Furthermore, the measurement beam is placed with a transversal offset and tilted around its center on the SEPD (Fig.~\ref{fig: off anschaulich}). 
According to Eq.~(\ref{eq: geometric}) there is no coupling for plane waves ($d_\text{pivot} = 0$), only the effect of the static offset remains.

For the analytical computation, we assume that the initial transversal offset changes the stationary reference beam $E_\text{ref}(x+d_\text{offset},y,z)$   instead of the measurement beam (since it is unimportant which beam is moved and the transformation of the measurement beam would be more complicated with an additional offset). 
The distance between pivot and SEPD is set to zero ($d_\text{pivot}=0$) and the measurement beam $E_\text{meas}$ is rotated around zero (see Eq.~(\ref{eq: rotation})). The expressions for the beams are the same as in Eq.~(\ref{eq: beam}).
The resulting overlap integral $O_\text{ovi}^{GO} $ for Gaussian beams with initial offset becomes:
\newline
\begin{widetext}
\begin{align}
O_\text{ovi}^{GO}&= \frac{ \exp\left(-\frac{k \left\{-i d_\text{offset}^2 \cos\alpha^2+(z_r-i z_0) \sin\alpha [2 d_\text{offset}+(z_0-i z_r) \sin\alpha]\right\}}{-z_0-3i z_r+(z_0-i z_r) \cos(2 \alpha)}\right) 2 \pi  \left(z_0^2+z_r^2\right)}{k \sqrt{z_r [-i z_0+3 z_r+(i z_0+z_r) \cos(2 \alpha)]}} \,.
\end{align}
\end{widetext}
This leads to the pathlength change:
\begin{equation}
s_\text{LPS} \approx \frac{-\alpha d_\text{offset}}{2} +\mathcal{O}(\alpha^2)\, .
\label{eq: offset coupling}
\end{equation}
This coupling is a result of a static offset. To compute the effect of the  dynamic (angle depending) offset in the initial case (as shown in Fig.~\ref{fig: geom anschaulich}), the offset itself ($d_\text{offset}$) has to be replaced by its geometric expression:
\begin{equation}
 d_\text{offset} = \tan(\alpha) d_\text{pivot} \approx \alpha  d_\text{pivot} \, .
 \label{eq: offset}
\end{equation}
By combining Eq.~(\ref{eq: offset coupling}) and (\ref{eq: offset})  the coupling caused by the offset in the initial setup becomes:
\newline
\begin{equation}
s_\text{LPS}\approx \frac{-\alpha d_\text{offset}}{2} = \frac{-\alpha^2}{2} d_\text{pivot} \, . \label{eq: result offset LPS}
\end{equation}

Therefore, the negligible tilt to $s_{\mathrm{LPS}}$ coupling of Eq.~(\ref{eq: phase to LPS}), is the result of
two effects:
The first one is an obvious geometric effect (Eq.~(\ref{eq: result plane wave})), which is the geometrical distance change between the pivot (beam origin) and the photodiode. 
The second one is a result of the offset between the two beams  which is also caused by the beam tilt. 
Both effects generate the same amount of coupling, but with a different sign. In the special case of two identical Gaussian beams on an infinite single element diode, the resulting coupling between beam tilt and measured pathlength  becomes negligible.

\section{Discussion}
\label{sec: discussion}
\begin{figure}
\centerline{\includegraphics[width=0.9\columnwidth , angle = {0}]{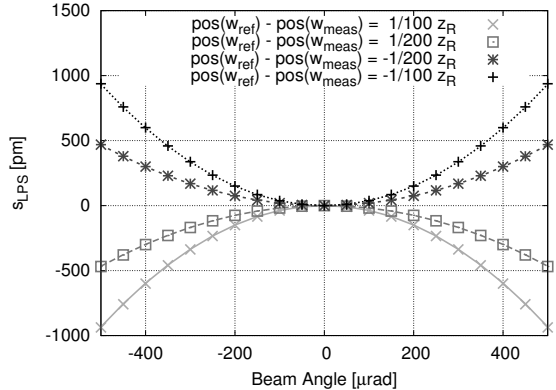}}
\caption{Numerically and analytically~\cite{Wanner2014} computed $s_{\mathrm{LPS}}$ for two Gaussian beams on an SEPD with differences in the waist position. The numeric values are marked with symbols, while the analytical expressions are lines. (Simulated setup is described in Sec.~\ref{sec: discussion}.)}
\label{fig: GG w}
\end{figure}

For unequal beam parameters ~\cite[Eq.~(34)]{Wanner2014} shows that additional coupling terms appear, which disturb the balance between both effects and lead to significant residual coupling. As an example, Fig.~\ref{fig: GG w} shows numerical simulations and analytical expressions of the coupling for the same situation as in Sec.~\ref{sec: gaussian beam} but with slightly different waist positions for the measurement and reference beam.
Similarly, incomplete detection, non-fundamental Gaussian beams and misalignment cause non-negligible tilt to length coupling, as we have observed in numerical simulations and more complex analytic computations.

Therefore, the effect described in this manuscript only appears under very specific circumstances. 
However, it can be used in various situations for example to stabilize an interferometer and investigate additional coupling effects.
A manuscript on experiments that make use of this effect to measure the tilt to length coupling caused by a quadrant photodiode is in preparation.

Another well known way to explain this effect for very small tilt angles is to express the tilt as an excitation of the Hermite-Gaussian (HG) 01  mode as explained in \cite{Anderson1984}. 
Due to the orthogonality of the HG modes this excitation will not change the pathlength readout when the entire interference pattern is detected. However, this is only an approximation that is valid for very small angles (much smaller than the far field divergence~\cite{Anderson1984}). The angles in the present examples exceed this limitation such that the field cannot be suitably expressed by an excitation of only the HG 01 mode.

When comparing the tilt to length coupling for plane waves Eq.~(\ref{eq: result plane wave}) to that of Gaussian beams Eq.~(\ref{eq: phase to LPS}) or Eq.~(\ref{eq: offset coupling}) it should be pointed out that Eq.~(\ref{eq: result plane wave}) is not a special case of Eq.~(\ref{eq: phase to LPS}) or Eq.~(\ref{eq: offset coupling}). Due to the assumed complete detection (infinite integration limits) the Gaussian beams can not be approximated by  plane waves.
 
\section{Conclusions}
\label{sec: result}
It was shown that the computed coupling between two plane waves matches exactly the expected geometric pathlength difference. In contrast, the coupling between beam tilt and longitudinal pathlength signal in an interferometer with two identical Gaussian beams and a large SEPD vanishes.
It was shown, that the reason for this disappearance is an additional coupling effect that is caused by lever arm between pivot and photodiode.
Building an interferometer with two identical Gaussian beams and a large SEPD is  a possible way to cancel out the tilt to pathlength coupling, for example in a homodyne interferometer with one light source.

\section*{Acknowledgement}
We gratefully acknowledge support by Deutsches Zentrum für Luft- und Raumfahrt (DLR) with funding of the Bundesministerium für Wirtschaft und Technologie with a decision of the Deutschen Bundestag (DLR project reference No 50 OQ 1301) and thank the Deutsche Forschungsgemeinschaft (DFG) for funding the Cluster of Excellence QUEST-Centre for Quantum Engineering and Spacetime Research.

\end{document}